%% file: main_last.tex
\documentclass[aps, prl, reprint]{revtex4-2}
\usepackage{graphicx} 
\usepackage{underscore} 
\usepackage{amsmath}

\begin{document}

\title{Electrostatic Phenomenology Benchmarks for Machine‑Learned Interatomic Potentials in Electrochemistry: Beyond the Energy‑Force Metric}

\author{Barbara Sumić}
\affiliation{Fritz-Haber-Institut der Max-Planck-Gesellschaft, 14195 Berlin, Germany}
\author{Ria Vasdev}
\affiliation{Department of Chemistry and Biochemistry, University of South Carolina, Columbia, SC, 29201, United States}
\author{Sudheesh Kumar Ethirajan}
\affiliation{Department of Chemical Engineering, University of California, Davis, Davis, CA, 95616, United States}
\author{Jing Yang}
\affiliation{Department of Materials, Imperial College London, London, UK}
\author{Clotilde S. Cucinotta}
\affiliation{Department of Chemistry, Imperial College London, London, UK}
\affiliation{Fritz-Haber-Institut der Max-Planck-Gesellschaft, 14195 Berlin, Germany}
\author{Richard G. Hennig}
\affiliation{Department of Materials Science and Engineering, University of Florida, Gainesville, Florida 32611, USA}
\author{Karsten Reuter}
\affiliation{Fritz-Haber-Institut der Max-Planck-Gesellschaft, 14195 Berlin, Germany}
\author{Stefan Ringe}
\affiliation{Department of Chemistry, Korea University, Seoul, Republic of Korea}
\author{Mira Todorova}
\author{Christoph Freysoldt}
\author{Jörg Neugebauer}
\affiliation{Max Planck Institute for Sustainable Materials, Max-Planck-Str. 1, 40237 D\"usseldorf, Germany}

 \begin{abstract}
          Accurate treatment of long-range interactions in machine learning interatomic potentials (MLIPs) is essential for electrochemical simulations. However, aggregate energy and force errors alone are insufficient to establish an MLIP's physical accuracy since they do not detect qualitative inconsistencies in the model such as the prediction of image-charge attraction, dielectric screening, or charge transfer. We introduce a benchmark suite EPhEct (Electrostatic Phenomena for Electrochemistry) of focused test cases designed to evaluate MLIPs on electrochemically relevant physical phenomena. The tests probe for image-charge attraction at a metal electrode, the splitting between longitudinal and transverse optical phonons as a probe of ionic and electronic screening, the dipole moment of interfacial water, and Fermi-level pinning during ion discharge. These tests establish a qualitative diagnostic routine complementary to aggregate energy-force metrics.

 \end{abstract}


\maketitle

\section{Introduction}

Accurate modeling of electrified reactive interfaces is a key part of the scientific workflow for understanding the mechanisms of a multitude of reactions directly relevant to the energy transition. Chemical processes such as charge-transfer, ion adsorption and desolvation, as well as reactive rearrangements, are decisive for efficiency and lifetime of batteries \cite{batteries_01, batteries_02, batteries_03, batteries_04}, fuel cells, electrolysers \cite{electrolyzers_01, fuel_cells_01} and supercapacitors \cite{supercapacitors_01, supercapacitors_02, supercapacitors_03, supercapacitors_04, supercapacitors_05, supercapacitors_06}, as well as degradation processes such as corrosion and formation of the solid-electrolyte interphase \cite{sei_01, sei_02, sei_03}. The bulk of the interfacial chemistry occurs within a nanometre-scale electrical double layer, across which the applied potential drops and the interfacial field is strongest \cite{edl_wu}.  The electric fields and voltages originate from the spatial separation of charges and are further modified by the screening response throughout the entire system; in turn, they affect the energetics of charged species and hence their distribution within the double-layer. Thus, the electrostatic effects are not only a methodological challenge to address, but form an essential target of simulations. What makes these systems demanding to model is that their behaviour is governed not only by the mean interfacial structure and energetics, but also by the fluctuations and rare events that modulate barriers and rates \cite{Magnussen2019,Todorova2026,Melander2023}. Specifically, the local electric field fluctuates due to the thermal motion of ions and solvent molecules; these fluctuations can significantly lower activation barriers for charge transfer -- a mechanism that is central to electrochemical reaction rates but difficult to capture with static mean-field approaches \cite{Surendralal2018PRL, smile-cp}.

Density-functional theory (DFT) remains the workhorse for approaching chemical accuracy at reactive electrified interfaces: it captures bond breaking and formation, chemisorption, and the voltage-dependent redistribution of electrons that ultimately controls screening and reactivity. Yet, DFT-based ab initio molecular dynamics (AIMD) is intrinsically confined to $\sim\!10^{2}$ atoms and $10$--$10^{2}$~ps of sampling with routine resources~\cite{Omranpour2024, Todorova2026}. Both bounds are far short of the regimes where solvent dynamics, electric double layer self-organization, interfacial morphology evolution, and rare events unfold. For example, water reorientation and hydrogen-bond
rearrangement, which gate charge transfer, occur on
$\sim\!1$--$10$~ps, whereas relaxation of the electric double layer,
which sets the electrode potential itself, requires
$\sim\!1$~ns~\cite{Melander2023}. An AIMD trajectory therefore samples
solvent reorganisation only marginally and never equilibrates the double
layer~\cite{Melander2023,Todorova2026}. Bridging this scale gap without giving up the electronic-structure accuracy that makes the chemistry meaningful is the main methodological challenge for predictive, operando modelling of electrochemical interfaces.

Machine-learned interatomic potentials (MLIPs) can reach nanosecond timescales for $10^{3}$--$10^{6}$ atoms at a per-step cost four to six orders of magnitude below DFT, while retaining near-DFT accuracy on forces~\cite{Omranpour2024,Jia2020}. The fundamental principles behind the development of MLIPs, pioneered by Behler and Parrinello in 2007 \cite{behler_parinello_2007}, are the nearsightedness of electronic matter \cite{kohn_1996} and the assumption that the total energy decomposes into atomic contributions, each depending only on the local environment within a cutoff radius. The potential energy surface of each atom can then be represented using a feed-forward neural network as a function of
features of the neighbouring atoms. These features are hand-designed, as in the original symmetry functions \cite{behler_parinello_2007} and in descriptor-based potentials such as SOAP-GAP \cite{gap_review, Bartok2013_SOAP} and ACE \cite{Drautz2019_ACE}, or learned, as in message-passing graph neural networks \cite{mace, nequip} that iteratively refine per-atom features from their neighbours. The effective range of the interactions is set either implicitly, by how far information propagates across the system (e.g. through message passing on a graph \cite{mace, nequip}), or explicitly, by augmenting the local model with long-range expressions (e.g. explicit electrostatics or long-range descriptors \cite{long_range_mlips_taxonomy}).

While the local approximation is highly successful for neutral bulk phases where long-range interactions are implicitly screened, electrochemical interfaces are among the most demanding targets for such models as the interface is charged, and screening effects extend over distances exceeding the MLIP cutoff. By construction, a strictly local model cannot capture the $1/r$ decay of electrostatic interactions beyond its cutoff, nor can it account for the non-local redistribution of charge required to describe metallic screening or dielectric polarisation.  
The underlying physics allows for a wide range of phenomena with different, dimensionality-dependent scaling behaviour when one moves from small-scale test and training cases toward the system sizes of interest for applications. A qualitatively correct and quantitatively accurate description, therefore, requires accounting for long-range Coulomb interactions, a dielectric and metallic response governed by the collective state of the electrode, and charge states that evolve as species approach, adsorb, or react. In practice, the implicit range of these common MLIP architectures extends only up to a few times the cutoff. While this captures some non-local phenomena \cite{kang_2024}, it remains insufficient for long-range electrostatic forces \cite{LES_cheng2025}. Non-local architectures \cite{SpookyNet_unke2021, AllScAIP_qu2026, SO3krates_frank2024}, on the other hand, can reach arbitrarily far, but scale less favourably with system size. Notably, neither class of methods encodes long-range electrostatics as an explicit physical term unless one is deliberately added. The local class cannot reach it, and the non-local class learns the interaction magnitude from data, with no inductive bias toward the correct $1/r$ asymptotic decay.  The extent to which such physical priors and inductive biases are necessary in MLIP development remains an active debate in the field \cite{6_questions_for_mlips, les_new_perspective, ReaxNet_gao2025, spice_v2}.

As highlighted in recent community discussions of MLIPs for electrochemistry \cite{6_questions_for_mlips}, long-range electrostatics, polarisation in the electrode environment, and charge transfer are recurring bottlenecks. A standard ``fit-to-energy/forces" metric cannot be decomposed into these physically distinct modes, meaning that two models with an identical total error can differ significantly in their ability to reproduce different phenomena such as image charge, charge state, or dielectric response. Furthermore, as demonstrated by Yang et al. \cite{smile-cp}, thermally induced electric field fluctuations at electrochemical interfaces are often dominant over the externally applied field. In such cases, standard fitting metrics may effectively describe the dominant 'noise' of the thermal fluctuations while failing to capture the weaker 'signal' of the applied potential response. Thus, a set of targeted diagnostics is needed to evaluate the performance of MLIPs on long-range and non-local phenomena. 

This work provides a phenomenological basis for evaluating and selecting MLIPs that have been proposed to capture long-range electrostatic effects. A general one-size-fits-all solution does not yet exist. To help evaluate MLIPs for specific scenarios, where the effects of interest can be narrowed down by the scientific question, we define a benchmark suite of test cases for electrostatic phenomena and provide DFT reference calculations; systematic evaluation of existing MLIPs against these tests is left for future work. The work grew out of the Fall 2025 program at the Institute for Pure and Applied Mathematics (IPAM) \cite{IPAM}: Bridging the Gap: Transitioning from Deterministic to Stochastic Interaction Modeling in Electrochemistry.

In that light, we introduce a benchmark suite EPhEct (\textbf{E}lectrostatic \textbf{Ph}enomena for \textbf{E}lectro\textbf{c}hemis\textbf{t}ry) containing four tests, each targeting a distinct electrostatic phenomenon that local MLIPs are prone to miss. All of the tests are carried out on systems relevant to electrochemical interfaces regularly studied by domain practitioners. The tests probe the image-charge effect (the attraction between a charge and the polarisation it induces in a metal surface), the splitting between longitudinal and transverse optical (LO/TO) phonons, the dipole moment of water, and Fermi-level pinning (the fixing of the Fermi level when band edges align and charge flows between them).

These phenomena stress MLIPs along four distinct axes. The first is range: whether the interaction energy decays correctly beyond the cutoff, which determines whether ion–ion, image-charge, or dipole–field interactions are captured. The second is non-locality: whether charge can redistribute across the system subject to global constraints and boundary conditions. These first two must be distinguished — an interaction can be long-ranged yet determined locally (fixed charges interacting through a $1/r$ tail), or short-ranged yet requiring non-local charge redistribution (charge equilibrating under a global constraint)~\cite{unke_2021}. The third is field response: whether the model reproduces the system's response to an applied field, which governs polarisation, capacitance, and the dielectric constant, and is itself both long-ranged and non-local. The fourth is charge state: whether the model represents how electronic charge redistributes as the chemical environment changes, since the same atomic geometry can correspond to different charge states, and a potential-energy surface that depends only on atomic positions cannot capture electronic redistribution that the positions do not encode.

Alongside the main test cases built on realistic systems, simple scaling tests are provided to test whether a model reproduces correct electrostatic behaviour in its simplest, most controlled form, as well as to enable extracting relevant physical quantities from energies alone, even when charges or potentials are not directly accessible. These tests establish necessary baseline behaviour -- a model that fails them cannot be trusted on the more complex interfacial systems that follow.

\section{Current methods}
\label{sec:CurrentMethods}

\begin{figure*}[t]
    \centering
    \includegraphics[width=\textwidth]{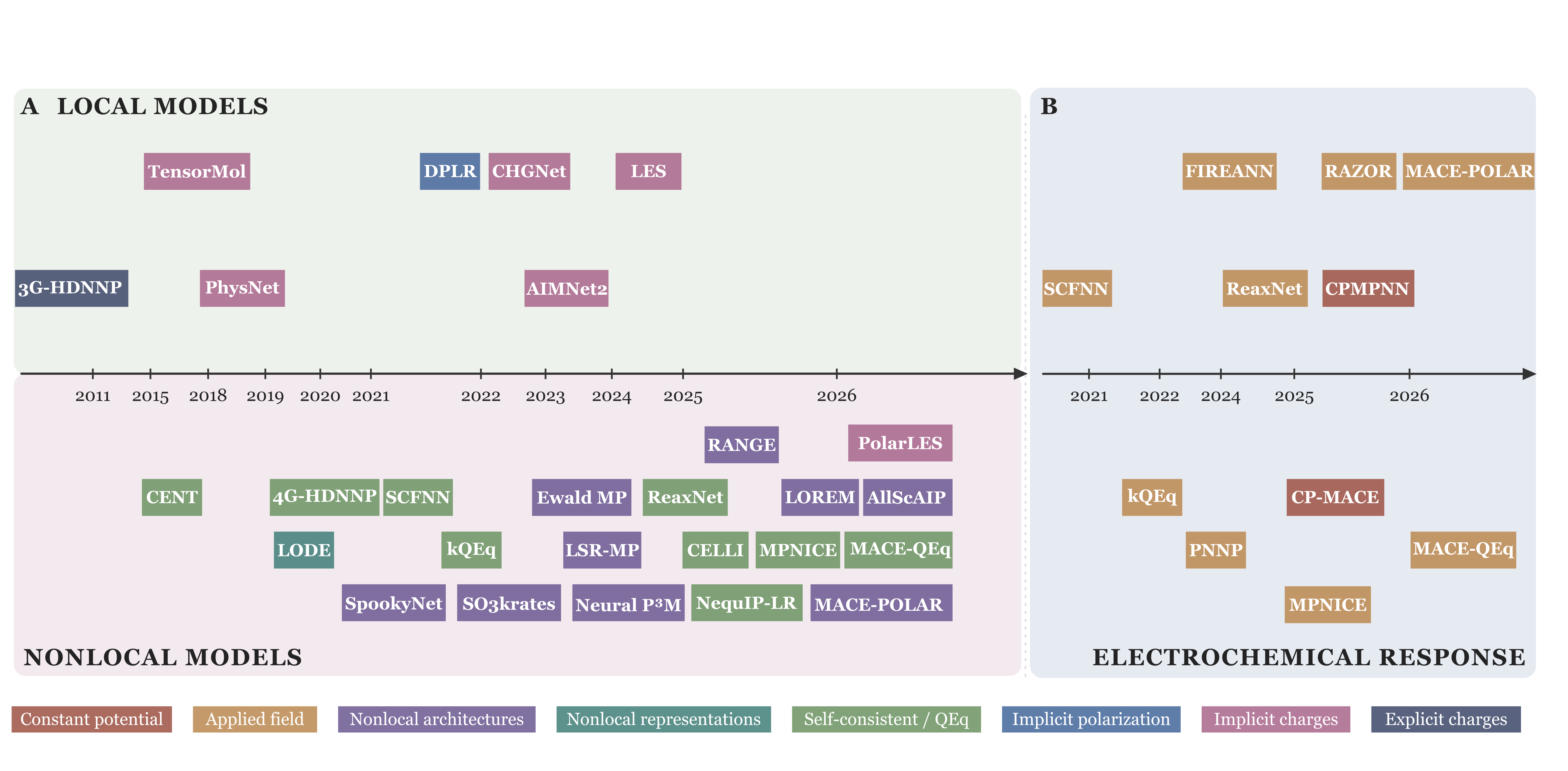}
    \caption{Timeline of MLIPs incorporating long-range electrostatics. \textbf{(A)}~Architectures are colored based on the taxonomy defined by Grasselli et~al.~\cite{long_range_mlips_taxonomy}: \emph{local models}---(i)~explicit charges, (ii)~implicit charges, (iii)~implicit polarisation; \emph{non-local models}---(iv)~self-consistent / QEq, (v)~non-local representations, (vi)~non-local architectures. \textbf{(B)}~Models capable of electrochemical response: applied-field models~(orange) accept an external electric field, constant-potential models~(red) take the electron number as input.}
  \label{fig:timeline}
\end{figure*}

Incorporating long-range electrostatics into MLIPs has taken several different paths (Fig.~\ref{fig:timeline}A).  Following the taxonomy of Ref.~\cite{long_range_mlips_taxonomy}, we group these into \emph{local} and \emph{non-local} models.  In local models, each atomic charge or feature is predicted from structural information within a cutoff radius. The resulting charges may still enter long-range Coulomb sums, but their values depend only on the local environment.  In non-local models, the learned predictions themselves depend on the global configuration, whether through self-consistent charge equilibration, system-wide descriptors, or architectural operations that couple distant atoms. 

Local models fall into three families: explicit charges, implicit charges, and implicit polarisation. \emph{Explicit charge models} train on DFT charge decompositions such as Hirshfeld partitioning and sum the resulting atomic charges via a Coulomb kernel~\cite{3G-HDNNP_artrith2011}. The charges are interpretable.  They are also partition scheme-dependent: no global neutrality constraint is enforced, and different partitioning schemes yield different MLIPs. \emph{Implicit charge models} sidestep this problem by treating charges as latent variables inferred from energies, forces, or dipoles. TensorMol~\cite{TensorMol_yao2018} and PhysNet~\cite{PhysNet_unke2019} took this route early, learning charges from local representations and computing explicit Coulomb interactions. The latent Ewald summation (LES)~\cite{LES_cheng2025, LES_king2025} predicts a hidden charge variable from local descriptors and computes its long-range interaction through the Ewald summation in electrostatic form, $1/r$. Although latent charges are learned purely from energies and forces, they recover physical dipole moments and Born effective charges~\cite{LES_king2025}, and the model reproduces correct dielectric screening at liquid-vapor interfaces~\cite{LES_cheng2025}. CHGNet~\cite{CHGNET_deng2023} uses DFT magnetic moments as a charge-proxy regularizer to pretrain a universal potential across the Materials Project, capturing charge-order phenomena such as lithium ordering in battery cathodes. AIMNet2~\cite{AIMNET2_anstine2025} embeds charge predictions directly into message-passing layers. \emph{Implicit polarisation models} learn Wannier-center positions from local environments, a representation grounded in the modern theory of polarisation~\cite{modern_polarization_theory_resta1992, modern_polarization_theory_king-smith1992, modern_polarization_theory_resta2007}. The longitudinal--transverse optical (LO--TO) phonon splitting is a direct signature of the macroscopic electric field in polar crystals. DPLR recovers this splitting in NaCl and achieves correct ionic screening at electrified TiO$_2$/electrolyte interfaces~\cite{DPLR_zhang2022, DPLR_edl_zhang2024}. However, non-local electronic polarisation is not guaranteed, as each Wannier-center position is predicted from the local environment alone~\cite{long_range_mlips_taxonomy}.

Non-local models likewise fall into three families: self-consistent, non-local representations, and non-local architectures. \emph{Self-consistent} charge equilibration (QEq) schemes optimize a charge-dependent energy under a global constraint $\sum_i q_i = Q$, so that each charge depends on the entire configuration~\cite{CENT_ghasemi2015, 4G-HDNNP_ko2020, kQEq_staacke2022, NequIP-LR_maruf2025, CELLI_fuchs2025}. Baldwin et al.~ \cite{self_consistent_design_space_baldwin2026} distinguish two QEq implementations. Energy-functional approaches minimize charge-dependent energy variationally, whereas fixed-point approaches iterate local charge predictions to convergence. However, when built on local ML models, the two give different results. The variational solution delocalizes the charge, favoring conductors, while the iterative solution better localizes the charge, favoring insulators, a distinction that matters at electrochemical interfaces where both phases coexist. SCFNN~\cite{SCFNN_gao2022} builds on Wannier centers but introduces a self-consistent loop between short- and long-range networks, making charge predictions depend on the global configuration. The model recovers the dielectric constant of bulk water with the accuracy of the underlying DFT functional. MPNICE~\cite{MPNICE_weber2025} performs QEq charge equilibration after each invariant message-passing iteration, feeding the equilibrated charges and long-range Coulomb interactions back into the next block at 5--20$\times$ lower cost than comparably accurate equivariant models. ReaxNet~\cite{ReaxNet_gao2025} extends QEq with environment-dependent polarisabilities inside an equivariant GNN, trained across the periodic table up to Pu as a broadly transferable foundation potential. These models have been tested on charge-state-dependent potential energy surfaces of NaCl and Ag clusters~\cite{4G-HDNNP_ko2020, QEq_test_cases_shaidu2024}, FeCl$_3$--water oxidation states~\cite{QEq_test_cases_kocer2025}, and water under applied fields, where QEq schemes can exhibit diffuse, unphysical polarisation~\cite{QEq_test_cases_kocer2025, MACE-QEq_vondrak2026}. \emph{Non-local representations} encode long-range information at the descriptor level: LODE constructs atom-centered features via a $1/r^p$ potential from all atoms and has been applied to electronic polarisation of metallic electrodes~\cite{LODE_grisafi2019, LODE-2_grisafi2023}. \emph{Non-local architectures} embed long-range operations directly inside the network. The mechanisms vary: reciprocal-space message passing~\cite{EwaldMP_kosmala2023}, mesh-based Fourier pathways~\cite{Neural-P3M_wang2024}, non-local or equivariant attention~\cite{SpookyNet_unke2021, SO3krates_frank2024}, equivariant long-range message passing~\cite{LOREM_rumiantsev2025}, fragment-level transformers~\cite{LSR-MP_li2023}, and virtual master-node attention~\cite{RANGE_caruso2026}. AllScAIP discards these structured long-range terms entirely, relying on data-driven all-to-all attention~\cite{AllScAIP_qu2026}. MACE-POLAR-1~\cite{MACE-Polar_batatia2026} combines learnable Fukui functions with a non-self-consistent polarisable field formalism built on Coulomb-kernel convolutions, embedding induction effects directly in the MACE architecture, and capturing non-local electronic polarisation without iterative self-consistency. The Polar LES~\cite{Polar-LES_kim2026} predicts latent monopoles, dipoles, and quadrupoles from local descriptors but applies non-self-consistent linear-response corrections via the global electrostatic field, recovering Born effective charge tensors, infrared spectra, and semi-quantitative Raman spectra for water and hybrid perovskites. These non-local approaches can capture arbitrary long-range correlations. However, learned interactions are not always directly interpretable as physical charges~\cite{long_range_mlips_taxonomy}.

\emph{Response learning} offers a complementary route: training on Born effective charges, polarisability tensors, or field-dependent potentials encodes the electric-field response without modifying the architecture~\cite{long_range_mlips_taxonomy}. FIREANN learns Born effective charges and polarisability tensors as energy derivatives~\cite{FIREANN_zhang2023}. PNNP learns atomic polar tensors from zero-field configurations alone and accurately predicts dielectric relaxation and field-dependent IR spectra of liquid water~\cite{PNNP_joll2024}. RAZOR extends this to electrified solid/liquid interfaces, learning work functions and Born effective charges to compute energy corrections up to second order at biased electrodes~\cite{RAZOR_bergmann2025}. Several QEq and non-local architectures from Panel~(A) can also operate under applied electric fields (Fig.~\ref{fig:timeline}B): kQEq, MPNICE, ReaxNet, and MACE-QEq incorporate the field via an $\mathbf{E}\cdot\mathbf{r}$ coupling in the charge equilibration step, SCFNN accepts it in its self-consistent loop, and MACE-POLAR includes it through its Coulomb convolution. CP-MACE takes a grand-canonical approach, accepting the electron number as input and predicting the Fermi level for constant-potential molecular dynamics at electrochemical interfaces~\cite{CP-MACE_wang2025}. CPMPNN redistributes a global excess-charge parameter through multihead attention over all atoms, achieving a three-orders-of-magnitude speedup over grand-canonical DFT for potential-dependent CO dimerization and hydrogen evolution barriers~\cite{CPMPNN_chen2025}.

\section{Scaling tests for correct electrostatic behaviour}

Before evaluating the four headline phenomena, it is instructive to define what constitutes correct electrostatic behaviour at a fundamental level. A robust description of electrochemical interfaces requires that a model satisfies four essential criteria: (1) entities such as individual molecules, ions, the liquid phase, or the surface -- must carry their intended charge; (2) interactions must exhibit the correct $1/r$ decay at long range; (3) the system must respond accurately to external fields and screening; and (4) charge redistribution must be physically constrained.

If the underlying ML model relies on an explicit electrostatic expression -- such as in latent charge or charge-equilibration methods -- and the implementation provides direct access to these charges, verifying these four criteria is straightforward. In such cases, checking the physical consistency of the charges should be considered an essential verification step. However, if electrostatic effects are only implicitly defined by the model architecture -- meaning the model provides only energies and forces without any explicit description of or access to atomic charges -- the following simple scaling tests offer a way to diagnose these behaviours indirectly.

\subsection{Single ions in periodic boundary conditions}
Since the electrostatic energy of truly charged, periodic systems would diverge, the energy expression
must account for an implicit homogeneous background. In addition, the energy is sensitive to shifts
in the electrostatic potential. Placing an ion in a cubic box with length $L$ with the average
electrostatic potential set to zero gives rise to
a scaling behaviour \cite{MakovPayne}
\begin{equation}
E(L) = E(\infty) - \alpha \frac{Q^2}{L} + C\frac{Q}{L^3}
\label{eq:Madelung}
\end{equation}
where $\alpha$=1.4186 in atomic units (Hartree$\cdot$bohr/$e^2$) is the Madelung constant
for a cubic lattice of point charges with
a neutralizing background and $C/L^3$ is the average electrostatic potential of the ion relative to that
of a point charge.
 Crucially, this diagnostic relies solely on the system energy, making it applicable even to models that do not provide explicit access to charges or electrostatic potentials. By fitting data from a series of box lengths to Eq.~(\ref{eq:Madelung}), one can extract the apparent charge $Q$ of the ion.

\subsection{Pair of ions in open boundary conditions} 
If the underlying electrostatic model does not support periodic boundary conditions, an obvious
test is the Coulomb law
\begin{equation}
E(d) = \frac{Q^2}{4\pi\epsilon_0 d} 
\label{eq:Coulomb}
\end{equation}
for a symmetric pair of ions at distance $d$. Again, fitting energy data against the distance with 
Eq.~(\ref{eq:Coulomb}) allows one to extract the charge value from energy data alone.

\subsection{Ions as probes for the potential} 
If the MLIP provides direct access to local charges, the corresponding electrostatic potential can be derived via the Poisson equation, and vice versa. However, even for ML models that do not provide explicit access to these quantities, the system energy alone is sufficient to probe the potential in the vacuum region and verify physically correct long-range behaviour. Once the charge value of an ion in vacuum has been verified, it can be used as a probe to measure the electrostatic potential in vacuum regions via energy differences. For this, one places
additional ions with opposite charges $Q$ and $Q'\approx -Q$ in the same position in two separate calculations, and determines
the potential via

\begin{equation}
V = \frac{\Delta E(Q) - \Delta E(Q')}{Q - Q'}.
\end{equation}
$\Delta E(Q) = E(\mathrm{system}+Q) - E_{\mathrm{iso}}(Q)$ is the
interaction energy of a test ion of charge $Q$ with the system, where $E_{\mathrm{iso}}$ is the energy of the isolated ion in the same simulation box as the system
of interest. The use of opposite charges minimizes the influence of the screening-induced
energy, that is proportional to $Q^2$.
The procedure assumes that no charge transfer occurs between the system of interest and the test ions; the combined-systems test below addresses this risk.

Similarly, the local field can be tested by a combination of two oppositely charged ions,
or a molecule with a known dipole (values of dipoles for charge-neutral molecules
can be extracted from their pair-wise interaction). For this, the orientation of the molecule
is flipped (or positions of the ion pair exchanged).

\subsection{Periodic slab systems}
The effective charge of a periodic slab system can be assessed using two distinct approaches, depending on the model's output capabilities. If the MLIP provides explicit access to the electrostatic potential, the effective charge can be determined directly by analyzing the vacuum potential profile. Alternatively, for models that provide only energies and forces, the charge can be inferred via the capacitor effect by examining how the total energy scales with the vacuum thickness. Given a slab of thickness $s$ with a charge $Q$ per area $A$, and a vacuum separation of $v$, the compensating background produces an effective countercharge primarily in the vacuum. The associated electrostatic potential profile $\overline V$, averaged parallel to the surface ($xy$ plane), exhibits a parabolic shape in the vacuum due to the homogeneous background
\begin{equation}
\overline V(z) = V_0 + V_1 z + \frac{Q}{2\epsilon_0 A (s+v)} z^2
\end{equation}
where $V_0$ (potential offset) and $V_1$ (electric field boundary condition) depend on the choice of origin.
By increasing the vacuum distance, the asymptotic behaviour in the total energy is given by (see SI)
\begin{equation}
E(v) = E_0 + \frac{Q^2 v}{24\epsilon_0 A}  + \mathcal O((s+v)^{-1})\;.
\end{equation}
$E_0$ is a fitting parameter that does not correspond to any meaningful limiting case.

Either the energy dependence upon changing the vacuum thickness, or the potential shape (in a single cell) can be used to extract the slab charge.

\subsection{Combined systems} 

Accurately capturing electronic screening and polarisation requires models to go beyond macroscopic averages, as the electronic response can vary drastically between different environments—for example, between bulk water and a water/metal interface. In such scenarios, local atomic charges may need to adjust to reflect these differences, while the total charge must remain conserved. For instance, if an ion moves parallel to an interface at a distance sufficient to suppress physical charge transfer, charge may redistribute within the interface or within the ion itself, but no unphysical charge transfer should occur between the two subsystems. Verifying this behaviour is critical before applying a model to electrochemical simulations.

The commonly employed charge equilibration (QEq) scheme \cite{kQEq_staacke2022,qpac,4G-HDNNP_ko2020,QEq_test_cases_shaidu2024,CELLI_fuchs2025,very-long-lr-mace} illustrates this challenge. In QEq, only the total charge is constrained, while the distribution is determined by minimizing the energy under the condition that the effective charging potential (i.e. $\partial E^{el}/\partial q_i$) is uniform across all atoms. When combining two subsystems with different intrinsic charging potentials, charge flows to equilibrate them. This effectively mimics a metallic system with a global Fermi level. While appropriate for metals, this behaviour is physically incorrect for semiconducting or insulating regions that possess a band gap near the Fermi level \cite{vondrak_2025}, and thus can sustain a potential difference without immediate charge flow. Consequently, QEq-based models are best suited for systems that are either metallic or experience only small potential differences, which limits their applicability in many electrochemical interfaces.

Similarly, models that predict local charges based solely on the local atomic environment \cite{LES_cheng2025} risk developing unphysical macroscopic space charges when constructing electrochemical interfaces. This can be avoided if either the training data ensures that the model learns global and local charge neutrality within suitable subsystems, or if the model architecture explicitly enforces this constraint. To mitigate such failures, training data should include configurations that introduce a spatial separation between potentially charged species, such as elongated simulation cells or asymmetric slab geometries with different configurations at the top and bottom side. Finally, testing for these artifacts -- either by directly inspecting local charges or indirectly by studying the energy scaling with vacuum separation -- can help identify incorrect charge distributions during the benchmarking stage.
\section{EP\lowercase{h}E\lowercase{ct} benchmark suite}

The four primary test cases expand on the foundational scaling tests to probe for each of the long-range limitations described in the Introduction: charge-state, range, non-locality, and field response. While the scaling tests are conducted on small systems and idealised geometries, the main test cases introduce additional complexity and move closer to realistic interfacial systems. Rather than simply asking whether electrostatics is encoded correctly at all, these tests are designed to determine if the model produces the correct physics on the systems practitioners actually study. This pairing helps distinguish whether a model's failure is fundamental (due to architectural limitations) or induced by complexity. The tests target qualitative reproduction, as a single scalar error metric cannot resolve different physical failure modes. We provide a DFT reference to establish a ground-truth answer, unambiguously defining whether a model ``passes" the test while making any failure qualitatively obvious. All four tests involve single-point calculations with no production MD or enhanced sampling. The only required outputs from the ML models are energies, forces, and their analytic derivatives. This makes the tests cheap, reproducible, and architecture-agnostic. Yet, the water dipole test
and the Fermi-level pinning test target the dipole and local charges, respectively; they are
convenient if these quantities are directly accessible. Otherwise, the field and potential probes
described in the previous section must be employed.
Consequently, while passing these tests confirms the presence of correct physical behaviour, it serves as a necessary but not sufficient condition for quantitative accuracy in dynamical simulations.

\subsection{1) Image-charge effect}

This test probes two of the four axes: the $1/z$ decay of the electrostatic interaction (range) and the redistribution of charge in the metallic electrode in response to the ion (non-locality). 

When a point charge $q$ is placed at distance $z$ above a planar conductor,
the electrostatic boundary condition at the surface is satisfied by a fictitious
image charge $-q$ located at distance $z$ below the surface. The resulting
interaction energy  is
\begin{equation}
    U(z)=-\frac{1}{4\pi\epsilon_0}\frac{q^2}{4z}
    \;,
    \label{eq:imageCharge}
\end{equation}
where the force is purely attractive.
The effect arises entirely from polarisation of
the surface electron density. The image effect is not restricted to metals, but also occurs at the surface of a dielectric, or at dielectric interfaces. In the dielectric case, the image charge and the resulting potential carry a factor $\frac{\epsilon_{B}-\epsilon_{A}}{\epsilon_A+\epsilon_B}=\frac{\epsilon_{A}^{-1}-\epsilon_{B}^{-1}}{\epsilon_A^{-1}+\epsilon_B^{-1}}$, where $\epsilon_A$ is the dielectric constant on the side containing the
real charge, and $\epsilon_B$ the one on the other side of the interface. The metallic case is recovered for $\epsilon_B^{-1}=0$.
\\\\
 The image effect is a fundamental component of electrified interfaces, e.g. when charged ions are in close proximity to a metallic slab. An ion traversing the electric double layer typically samples separations of 5--20~\AA, where the $-1/z$ tail is most consequential.

A range of interfacial properties is affected by this interaction. Missing the image charge interaction artificially reduces the predicted interfacial capacitance, a critical quantity in supercapacitors and ionic-liquid interfaces.
Classical MD results show that under an external field, surface polarisation strongly enhances the charge separation across the electrolyte; consequently, the capacitance of the cell is greatly enhanced for a system with a conductive surface, as opposed to a low-dielectric (non-polarisable) surface \cite{son_image-charge_2021}. An MLIP that omits the image charge interaction effectively treats the electrode as non-polarisable, therefore underestimating the capacitance. 
Furthermore, the image charge attraction stabilizes the adsorbed ions. Although the magnitude of this interaction is heavily attenuated by the solvent screening -- ranging from chemical bond-like in vacuum to intermolecular bond-like in aqueous solution \cite{geada_insight_2018} -- even an intermolecular-bond-scale contribution is several $k_{\mathrm{B}} T$. This represents a meaningful systematic bias to the energy captured by an MLIP.
Finally, the image charge interaction at a metal surface enhances the charge-separation fluctuations across the electrolyte, which in turn affect the charging and relaxation dynamics. The lack of image effect may lead to inaccurate kinetic barriers and residence times \cite{son_image-charge_2021}.

\begin{figure}[th] 
\centering
\includegraphics[width=0.45\textwidth]{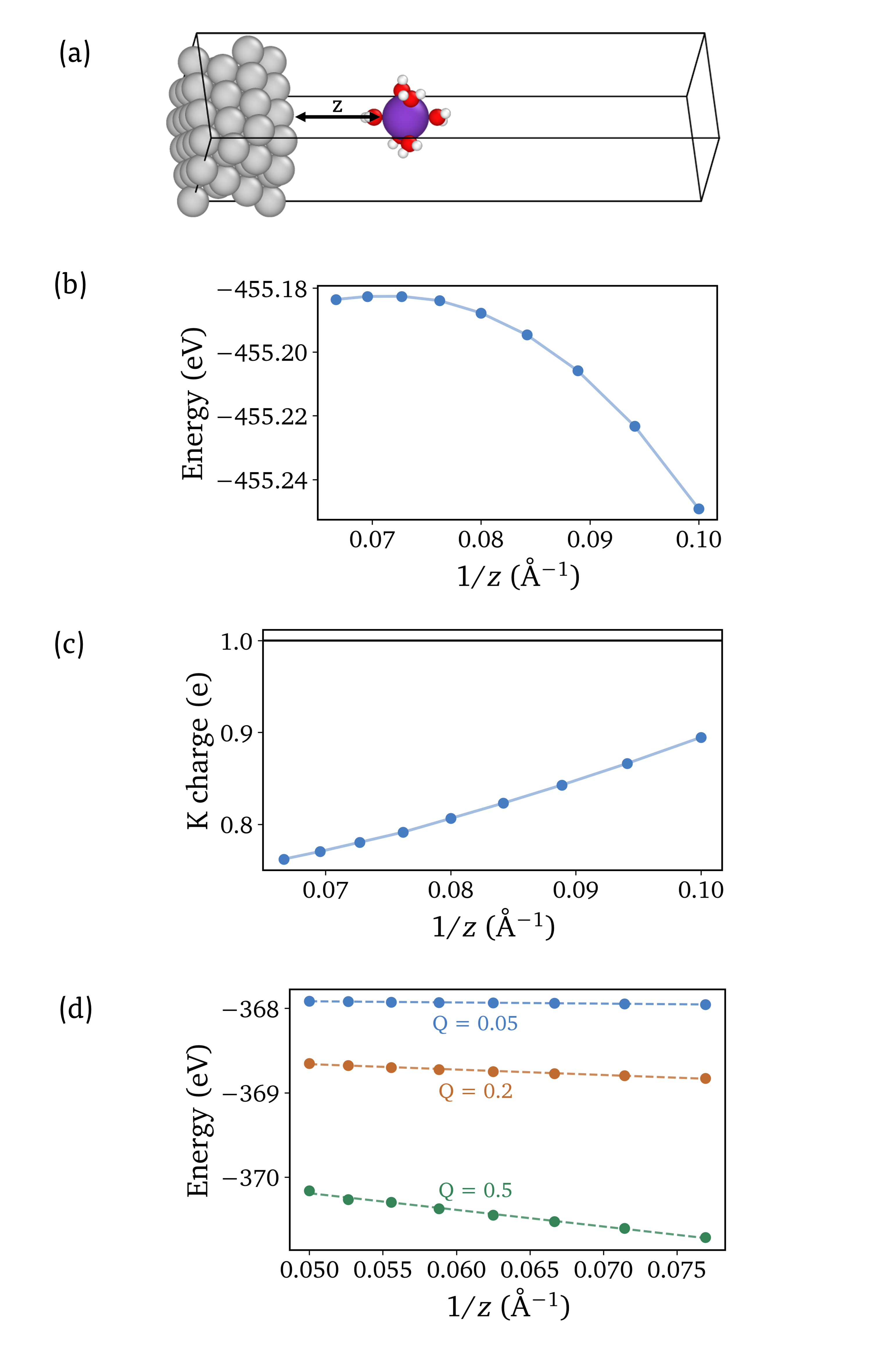}
\caption{\label{fig:image_charge} (a) The atomistic model of a Pt (111) surface with a K ion with its first solvation shell of water placed at distance $z$ from it. (b) The total energy of the system as a function of 1/$z$. (c) The net charge on the K-water complex as a function of 1/$z$. (d) The total energy of the system replacing the K ion with a probe charge $Q$, showing a linear 1/$z$ dependence. }
\end{figure}
The nearsightedness of standard MLIPs is fundamentally incompatible with the image interaction for several reasons. First, the $-1/z$ potential is long-ranged and algebraically
decaying, remaining significant beyond any practical cutoff. Second, it is
non-local: its magnitude depends on the collective dielectric and metallic
response of the entire electrode, not only on the atomic environment within the
cutoff sphere. Third, it is charge-dependent; a neutral atom and an ion at the same position interact very differently with the surface, which cannot be distinguished by most MLIPs as they do not carry the concept of explicit ionic charge.

In Fig. \ref{fig:image_charge}, we show an example of explicit DFT calculation, with a K ion with its first solvation shell of water placed at different distances towards a Pt(111) surface. The total energy as a function of $1/z$ is shown in Fig. \ref{fig:image_charge}b. It should be noted that in such calculations, it is difficult to observe a pure image charge effect as in Eq. \ref{eq:imageCharge} for several reasons. First, Eq. \ref{eq:imageCharge} applies to a semi-infinite metal plate. In the simulation, periodic boundary condition is applied to the $xy$ direction and the energy is dependent on the lateral size. Second, it is difficult to localize a constant +1 charge on the K atom due to Fermi level pinning (see discussions also in Test case 4). Fig. \ref{fig:image_charge}c shows the K charge state obtained as a function of $1/z$, clearly showing a distance dependence. Third, in consequence of the partial ionization of K, part of the +1 charge will sit at the metal surface, introducing an additional energy contribution of the charged surface interacting with the ion. As a result, the energy deviates from a linear $1/z$ dependence. 

Alternatively, one can insert a probe charge in replacement of the K atom (see details in SI). Fig. \ref{fig:image_charge}d shows the energy dependence as a function of $1/z$ with a probe charge $Q$. In this case, we get rid of the partial ionization issue and retrieve a linear $E_{\mathrm{tot}}$-$1/z$ dependence. However, given the finite-size effect, the energy dependence doesn't strictly follow Eq. \ref{eq:imageCharge}.

\subsection{2) Ionic and electronic screening}
This test probes the field-response axis: whether an MLIP distinguishes electronic from ionic screening, using the LO--TO phonon splitting as the observable. Polarisation and screening in a material arise from the response of the electrons to the effective electric field (electronic screening) and from the displacement of the ions, which generates a polarisation. In this view, screening from dipolar orientation, ionic displacement, piezo-electric effects, etc. are all classified as ionic polarisation.
Ionic polarisation must be captured by any electrostatically augmented MLIP. As for the electronic screening, the polarisability may be a direct function of the atomic environment and thus naturally integrated into an interatomic potential framework, while the electronic polarisation is generally not uniquely defined by the atomic environment.

In a single isotropic material, the electronic long-range screening effects can be effectively described by the so-called \textit{high-frequency} dielectric constant $\epsilon_\infty$.
In anisotropic materials the directional dependence of the screening can be captured by a dielectric tensor. This anisotropy stems from an uneven orientational distribution of the polarisable inter-atomic bonds across the system and is thus an emergent property of the local bonding environments. Consequently, treating it as a global parameter in a model, independent of these environments, is unphysical. We will therefore restrict ourselves to isotropic screening in the following.
The Coulomb interactions between any two charges $Q$, $Q'$ at sufficiently large distance inside such an isotropic material depend on $QQ'/\epsilon_\infty$, and it is tempting (and commonly done \cite{Jorge2024,Zhong2025}) to include the electronic screening in rescaled charges $Q_{\rm eff}=Q/\sqrt{\epsilon_\infty}$.

Yet, doing so introduces additional complexity in the coupling to external electric fields and strain.
In the original formulation with explicit electronic screening, displacing an atom produces
an (unscreened) polarisation $\Delta \mu_x = Q\Delta x$, which will be screened inside the material to produce
a field of $\mathcal E_x=\frac{1}{\epsilon_0\epsilon_\infty}\Delta\mu_x/\Omega$ in a cell of volume $\Omega$. The unscreened
polarisation appears, however, if the system is finite in the direction of the polarisation. Applying an external field
$\mathcal E^{\rm ext}$ to such a finite system interacts with the unscreened polarisation, i.e., $\Delta E = -\mathcal E^{\rm ext}_x\Delta \mu_x$. Conversely, such an external field will produce an extra force for an atom inside the material
of $\frac{1}{\epsilon_\infty}\mathcal E^{\rm ext} Q$. To take this into account with rescaled charges, the external field needs to be rescaled by an unphysical factor $\frac{1}{\sqrt\epsilon_{\rm \infty}}$, treating internal and external fields no longer on equal footing.

In the limiting case of a metallic screening (1/$\epsilon_\infty\rightarrow 0$), the ionic charges are
no longer able to produce electric fields inside the metals.
However, surface charges may still exist.

In ionic dielectric materials, the ratio between the high-frequency dielectric constant, $\epsilon_\infty$ and \textit{the static dielectric constant} $\epsilon_{\mathrm{s}}$ (which includes the ionic screening) is linked to the splitting between longitudinal optical (LO) and transverse optical (TO) phonon mode frequencies $\omega$ near the $\Gamma$ point via the
Lyddane-Sachs-Teller ('LST') relation \cite{LyddaneSachsTeller}. In a diatomic, isotropic system such as MgO it reads
\begin{equation}
\frac{\epsilon_{\mathrm{s}}}{\epsilon_\infty} = \frac{\omega_{\mathrm{LO}}^2}{\omega^2_{\mathrm{TO}}} = \frac{D_\mathrm{LO}}{D_{\mathrm{TO}}}\;,
\end{equation}
where $D_{\mathrm{LO/TO}}$ denotes the phonon mode force constant. Generalized LST relations hold for polyatomic and anisotropic cases \cite{ChavesPorto}.

\begin{figure}
\includegraphics[width=0.9\columnwidth]{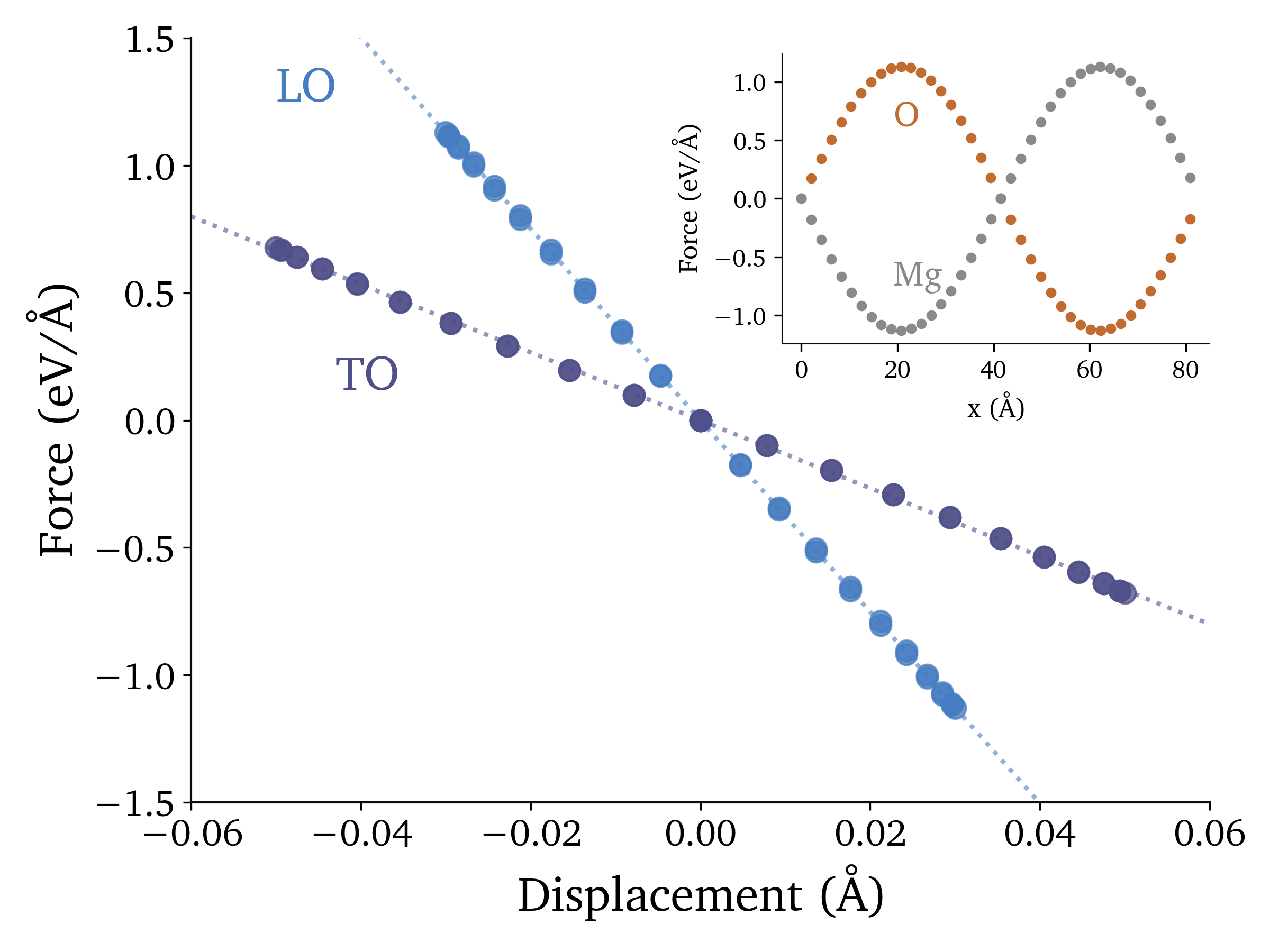}
\caption{Forces from frozen-phonon modes in a $(20\times 1 \times 1)$ MgO supercell. The inset shows the forces along the long cell axis. The slope of LO and TO modes correspond to the respective force constant. \label{fig:MgO_LO_TO}}
\end{figure}

The LO/TO benchmark for dielectric electronic screening works by setting up finite phonon modes at finite wavevector in elongated cells (e.g. along $x$ with length $L$), and displacing the ions either in the direction of the long dimension (LO) or perpendicular to it (TO). Displacing each atom (type $s$ located at $x_0$) by $C_s \sin (2\pi x_0/L)$ with the site-specific amplitude $C_s$ given by the optical phonon eigenmode (e.g. equal amplitude and opposite direction in a diatomic
system) produces forces proportional to the displacement, from which the effective force constant $D_{\mathrm{LO/TO}}$
can be obtained. Fig.~\ref{fig:MgO_LO_TO} illustrates the procedure. The existence of the LO-TO phonon splitting is
thus a sensitive test for a correct qualitative dielectric behaviour. A quantitative agreement of the LO/TO force
constant ratio with reference data ensures a correct asymptotic behaviour in homogeneous setups. However, it does not allow
extracting the electronic and static dielectric constant separately.
The dipole test in a slab geometry, that is described below for water, gives access to unscreened dipoles, and then also (Born) charges $Q_i=\frac{d\mu_z}{dz_i}$ upon displacing a single ion $i$.

Challenges arise in heterogeneous systems in which the (local) high-frequency dielectric constant varies. The effective screening
between any two points in this case cannot be reduced to a function of the local dielectric constants, and rescaling the charges is
therefore bound to fail.
An electrostatically consistent treatment
would need to include some kind of (electronic)
local polarisability  $\alpha_i$ (per atom $i$) or dielectric constant $\epsilon_\infty(\mathbf r)$ (per position). To account for the system geometry, one
must obtain a self-consistent solution of the
polarisation equation
\begin{equation}
    \boldsymbol{\mu}^{\rm el}_i = -\alpha_i \nabla V(\mathbf r_i)
\end{equation}
for each atom $i$ where the electrostatic potential $V$ arises from the ionic charges and the electronic dipoles,
or the locally screened Poisson equation
\begin{equation}
    \nabla \cdot \left\{\epsilon_0 \, \epsilon_{\infty}(\mathbf r)\nabla V(\mathbf r)\right\} = -\rho(\mathbf r)
\;.
\end{equation}
Due to the superposition principle in electrostatics, both equations are linear in their parameters, and
efficient solvers for the relevant algebraic equations (for $\boldsymbol{\mu}^{\rm el}_i$) or partial differential equation, respectively, are available.  To the best of our knowledge, such a treatment for dielectric systems has not yet been combined with MLIPs.

\subsection{3) Water dipole in a slab geometry}

The dipole moment serves as an important metric for evaluating whether a given MLIP predicts the correct electrostatic behaviour. 
It is a vector quantity with three components ($x$, $y$, $z$) that can be readily incorporated into the existing MLIP architectures and captures the global electrostatic behaviour. The charge density and the electrostatic potential, by contrast, are three-dimensional continuous fields whose direct incorporation into a loss function is non-trivial.
This metric has been applied in the development of universal MLIPs targeting molecular chemistry \cite{MACE-Polar_batatia2026, self_consistent_design_space_baldwin2026, Gonnheimer2026}, using existing datasets of molecular dipoles \cite{Hait2018JCTC,Liang2025JCTC}. Here, we focus on the modeling of the electrochemical interface and discuss how dipole moments can be used as a benchmarking metric. We limit the discussion to solid/water systems but the principles apply to other molecular liquids as well.
The description in the following assumes that the MLIP provides direct access to
potentials and/or dipoles. The auxiliary setup for energy/force-only models is described in the SI.

\begin{figure}[th] 
\centering
\includegraphics[width=0.5\textwidth]{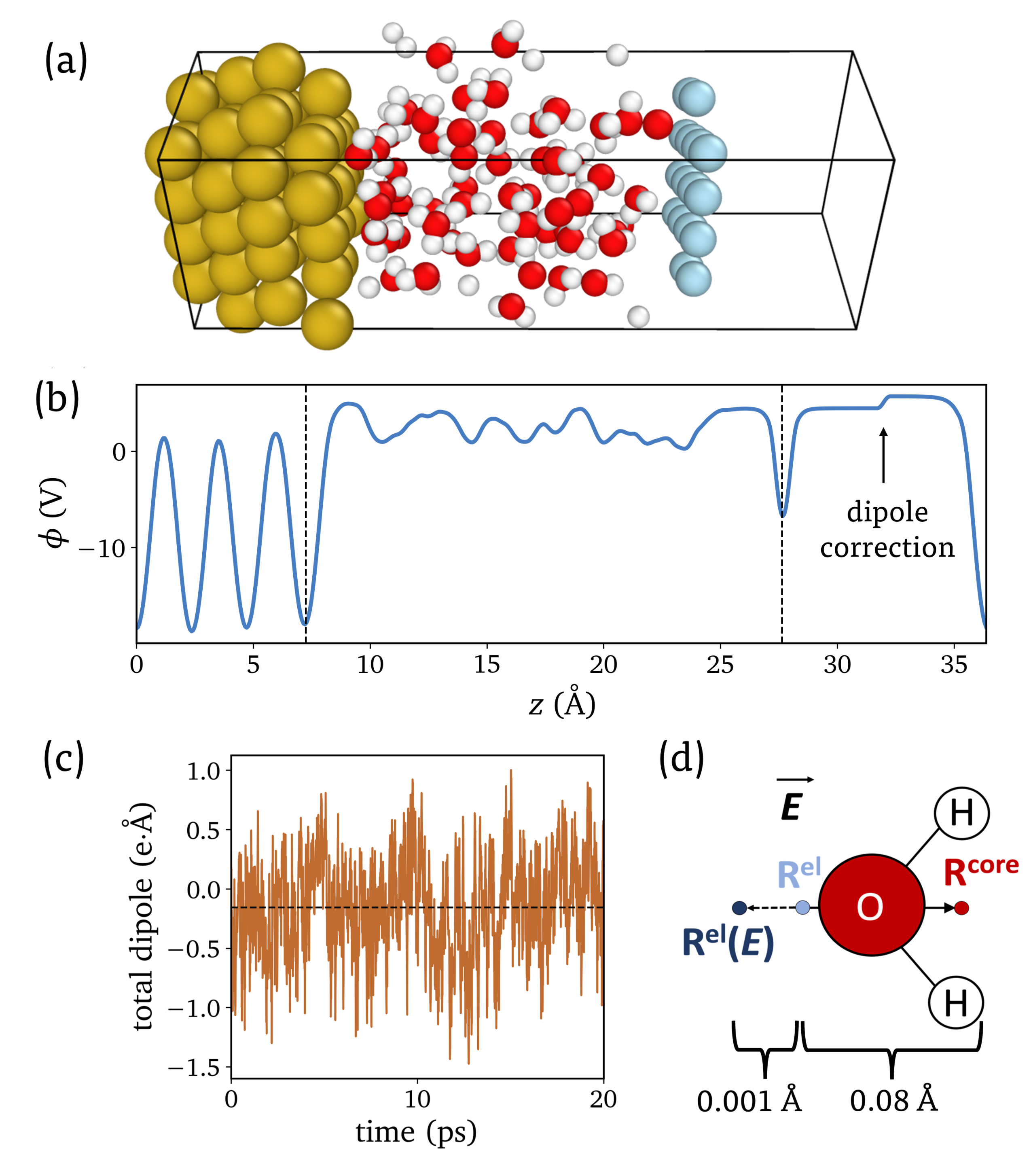}
\caption{\label{fig:dipole} (a) The atomistic model of a gold (111)/water interface system. The blue spheres represent the Ne counter electrode. (b) The corresponding electrostatic potential $\phi$ of a given frame, averaged on the $xy$ plane and plotted as a function of the $z$ (surface normal) direction. The dashed black vertical lines represent the position of the gold surface and the Ne counter electrode, respectively. The black arrow marks the dipole correction. (c) Evolution of the total dipole moment in the $z$ direction of the system during a 20 ps ab initio molecular dynamics simulation. The horizontal black dashed line marks the averaged value. (d) Schematics of how the individual water molecule dipole gets polarised under electric field (see discussion in text). Figure adapted from Ref. \cite{smile-cp}.}
\end{figure}

Unlike for single molecules, obtaining the dipole moment of a periodic system is less straightforward due to its multi-valued nature \cite{Spaldin2012}. For solid/water systems, there are two commonly-used approaches to obtain the dipole moment. First, one can break the periodicity in the surface normal ($z$) direction. A typical simulation setup is shown in Fig. \ref{fig:dipole}a, for the model system of a gold (111)/water interface. The computational counter electrode \cite{Surendralal2018PRL} is added on the back side but the following discussion applies also to cases without the counter electrode. In this setup, the top of the water slab is separated from the back side of the gold layer by vacuum. A dipole correction is then applied in the vacuum region to remove the spurious
electrostatic interactions between periodic images in the $z$ direction \cite{Neugebauer1992}.   

By breaking the periodicity in the $z$ direction, the total dipole in the $z$ direction $\mu_{\mathrm{tot}}$ (the subscript $z$ is omitted for simplicity) can be easily calculated by integrating the charge density. In Fig. \ref{fig:dipole}b, we show the electrostatic potential $\phi$ profile along the $z$ axis for a given frame. The step height at the dipole correction point $\Delta\phi$ is connected to  $\mu_{\mathrm{tot}}$ by
\begin{equation}
\Delta \phi = \frac{\mu_{\mathrm{tot}}}{\epsilon_0 A},
\end{equation}
where $\epsilon_0$ is the vacuum permittivity and $A$ is the surface area in the $xy$ plane. In this way, $\mu_{\mathrm{tot}}$ serves as a measure of the long-range electrostatic behaviour of the system. Due to the large thermal fluctuation of the liquid electrolyte, $\mu_{\mathrm{tot}}$ fluctuates on the order of 1\,$e\cdot$\AA{} during an AIMD simulation. Fig.~\ref{fig:dipole}c shows the evolution of $\mu_{\mathrm{tot}}$ over a 20~ps trajectory. 

The second approach to obtain the water dipole moments is to use the maximally localized Wannier functions (MLWFs) \cite{marzari1997maximally,marzari2012maximally}. The center of the MLWFs, also called Wannier centers (WCs), offers a way to decompose the charge density into a distribution of point charges. Under this description, each water molecule is associated with four WCs, and each WC with two electrons. For each water molecule, one can then calculate the center of mass of the nuclear core charge $\mathbf{R}^\mathrm{core}$ and the electron charge $\mathbf{R}^\mathrm{el}$. The individual water molecule dipole can then be calculated by
\begin{equation}
    \boldsymbol{\mu}_{\mathrm{H_2O}} = 8e(\mathbf{R}^\mathrm{core} - \mathbf{R}^\mathrm{el}). 
\end{equation}

A key consideration for MLIP development is that accurate reproduction of $\boldsymbol{\mu}_{\mathrm{H_2O}}$ does not guarantee a correct description of $\mu_{\mathrm{tot}}$, and vice versa. Ref.~\cite{smile-cp} discusses this mismatch; we summarize the key observations here. MLIP models that predict WC positions, such as the deep potential long range model (DPLR) \cite{DPLR_zhang2022}, have an accuracy of the order 0.005 $\mathrm{\AA}$. The average displacement between $\mathbf{R}^\mathrm{el}$ and $\mathbf{R}^\mathrm{core}$ is about 0.08 $\mathrm{\AA}$ so the DPLR model captures the individual water dipole within 5\% error. However, when the predicted $\boldsymbol{\mu}_{\mathrm{H_2O}}$ are summed up and projected to the $z$ axis, they deviate systematically from $\mu_{\mathrm{tot}}$. 

This seemingly contradictory observation relates to the electronic polarisation of water, which is illustrated schematically in Fig. \ref{fig:dipole}d. When a water molecule is placed in an electric field, its electron density becomes polarised, leading to the displacement of $\mathbf{R}^\mathrm{el}$.  For an electric field of 0.2 V/\AA, the displacement of $\mathbf{R}^\mathrm{el}$, $\Delta\mathbf{R}^\mathrm{el}$, is on the order of 0.001\,\AA. Looking only at $\boldsymbol{\mu}_{\mathrm{H_2O}}$, $\Delta\mathbf{R}^\mathrm{el}$ is too small to manifest in the model accuracy. However, when a long-range electric field exists throughout the system, which typically includes several tens of water molecules, the collective electronic polarisation contribution shifts $\mu_{\mathrm{tot}}$ by around 0.5 $e\cdot$\AA. (This value is half of the $\mu_{\mathrm{tot}}$ magnitude, which is connected to the electronic dielectric screening constant of water $\epsilon_{\infty} \approx 2$.) In other words, MLIP models that learn from local charge decomposition fail to capture the signal of the long-range electric field response, leading to a systematic error in the total dipole prediction. 

Similarly, if a model learns naively on $\mu_{\mathrm{tot}}$, it fails to assign a collective $\Delta\mathbf{R}^\mathrm{el}$ to each water molecule (see the SMILE0 model in Ref. \cite{smile-cp}). Instead, the model reduces the static dipole of each water molecule by half to compensate for the absence of electronic polarisation. Such model failure can be detected by looking at the individual water dipole distribution. An underestimation of the individual water dipole magnitude also leads to weaker screening of water. The weakened screening can be tested by inserting a dissolved ion into water and inspecting the electrostatic potential around it. 


It has also recently been shown that short- and long-range MLIPs behave qualitatively differently in predicting the macroscopic dipole distribution, even when they predict near-identical atom density profiles at a metal/water interface \cite{parker2026falsemetallizationshortrangedmachine}. The short-range MLIP falsely produces frames with very large macroscopic dipoles, which cause dielectric breakdown in the DFT calculations. This example again shows that the global dipole is an informative metric for evaluating the performance of MLIPs in long-range electrostatics. 

In summary, the dipole moment offers rich information on the electrostatic behaviour of water. Although obtaining the dipole moment of a periodic system is not straightforward, methods exist to calculate both the global dipole and the individual molecular/atomic dipole. An MLIP model that aims at describing the electrostatics at a solid/water interface should be benchmarked against both the individual water molecule dipole and the global system dipole. A model that accurately predicts both ensures the correct reproduction of the electrostatic potential profile \cite{smile-cp}.

\subsection{4) Fermi-level pinning}

Dielectric breakdown and Fermi-level pinning are particular manifestations of long-range electrostatic effects. Both are related to the relative position of the band edges of the material(s) under consideration and the redistribution of electrons as levels align. 

We first discuss dielectric breakdown. In condensed matter phases with an electronic gap, such as semiconductors or water, the presence of a surface or interface tilts the band edges due to differences in the electronic structure across the boundary. This effect is particularly pronounced at the interface of two different condensed phases in contact with each other and results in charge rearrangements in the vicinity of the interface, which aim at restoring the bulk level positions at some distance from the interface. In semiconductors, the charge rearrangement region is conventionally referred to as a space-charge region, in electrochemistry as the electric double layer. In the presence of an externally applied electric field these effects are enhanced and may lead to situations where the band edge of the occupied states on one side of the interface dips below the band edge of the unoccupied states on the other side of the interface. The band crossing invokes a charge transfer/redistribution that aims at aligning these bands to the same energy level, which renders the system metallic, fixes the Fermi level and thereby effectively prevents any further increase of the externally applied field (for more in-depth discussions see Refs. \cite{yoo2021, Todorova2026}) or changes in charging. Since this behaviour has practical consequences for calculations of electrochemical interfaces, it needs to be captured by MLIPs to ensure a faithful description.  

In the context of electrochemical interfaces, such alignment issues become particularly important when studying the impact an applied potential has on reactions \cite{bergmann_jcp}. Consider the type of system predominantly studied in the ab initio electrochemical literature -- a metal surface in contact with water. The setup is typically modeled by a metal slab in contact with a slab of neat water.  When a field is applied, the band edges of water will tilt and the Fermi energy of the metal, which normally lies somewhere within the band gap of water, may, depending on magnitude and the direction of the field, either dip below the valence band maximum of water or rise above its conduction band minimum.

\begin{figure}[th] 
\centering
\includegraphics[width=0.5\textwidth]{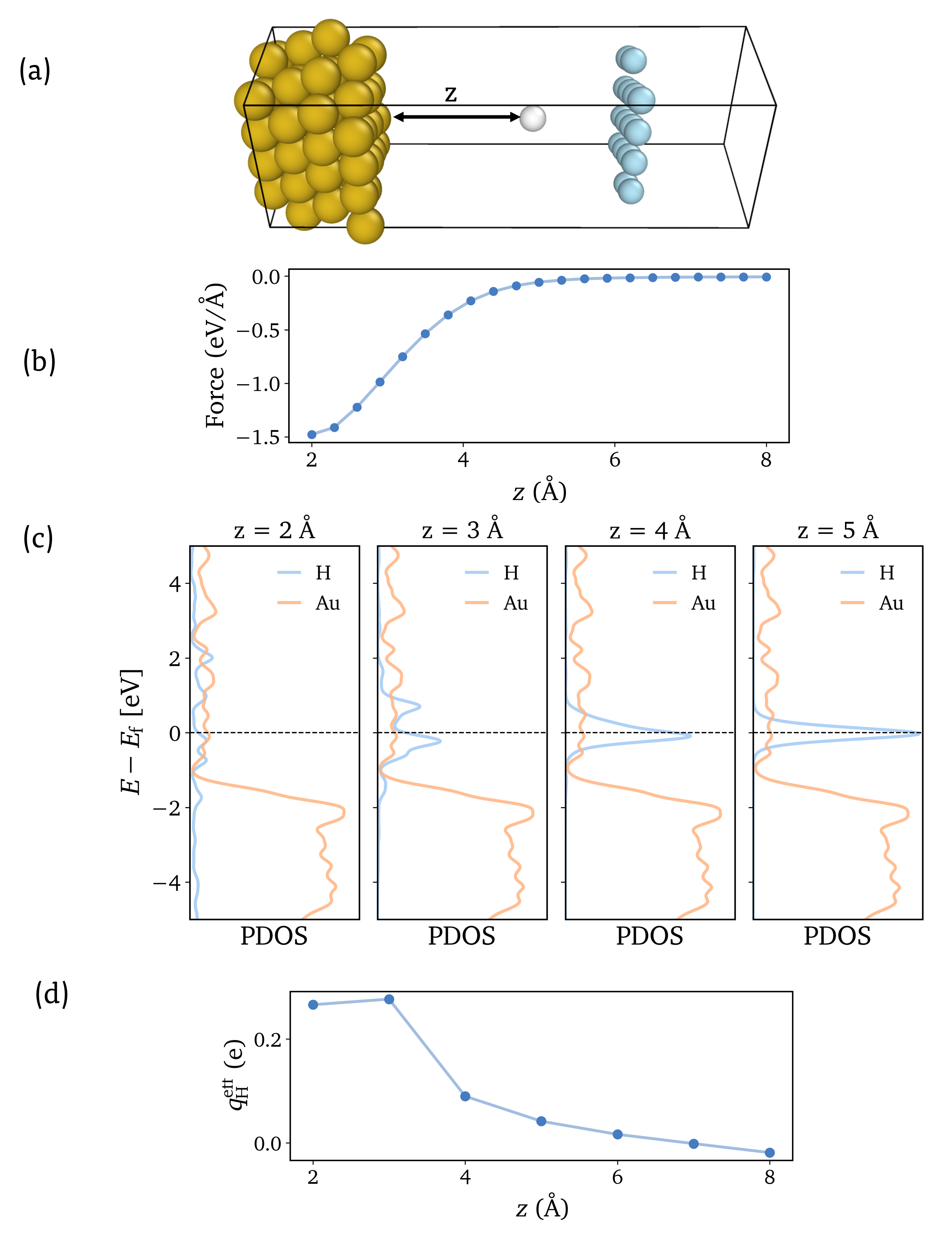}
\caption{\label{fig:Hads} (a) The atomistic model of a gold (111) surface with one hydrogen atom placed at different distances $z$ away from the surface. $z=0$ is defined by the outermost Au layer. (b) The force on the hydrogen atom as a function of $z$. (c) The partial density of states (PDOS) of gold and hydrogen at different $z$. For visual clarity, the PDOS of hydrogen is scaled by a factor of 100. (d) The effective charge of the hydrogen atom as a function of $z$. }
\label{FermiPinning}
\end{figure}

Another case where charge redistribution and Fermi-level pinning become important, and which presents a simple test case for MLIPs, is shown in Fig.~\ref{FermiPinning}. There, a H atom is moved from an adsorbed position on a Au(111) surface to a position within the vacuum region away from the surface (Fig.~\ref{FermiPinning}\,a). The density of states shown for different positions of H above the surface (Fig.~\ref{FermiPinning}\,c) and the corresponding charge of H (Fig.~\ref{FermiPinning}\,d) depict the expected charge transfer, which eventually renders the H neutral when it no longer interacts with the surface. 

With an MLIP it would not be possible to directly observe the discharge of an ion as it moves away from a surface (or the transition from an atom to an ion, as it dissolves from a metal surface in contact with water) by plotting either the density of states for the system or the position dependent charge on the particle. However, it is possible to probe the charge state through the force on the particle. A homogeneous applied field $E_{\rm field}$ exerts the additional force $q^{\rm eff}E_{\rm field}$ on a particle of effective charge $q^{\rm eff}$; the field-induced force therefore remains finite while the particle stays charged and vanishes once the particle becomes neutral. Tracking the field-induced force as a function of the distance from the surface thus maps out the charge profile $q^{\rm eff}(z)$ shown in Fig.~\ref{FermiPinning}\,d (the SI describes the corresponding procedure).

\section{Conclusion}

Energy and force errors, the standard metrics of MLIP quality, cannot resolve whether a model captures the electrostatics that govern electrochemical interfaces. We have therefore proposed EPhEct, a suite of four tests that target qualitative correctness -- the image-charge effect, ionic and electronic screening probed by LO-TO phonon splitting, the dipole moment of interfacial water, and Fermi-level pinning -- each isolating a distinct long-range or non-local failure mode against a DFT reference, complemented by scaling tests that verify correct electrostatic behaviour in its simplest form.

Such qualitative diagnostics reveal distinctions that aggregate errors hide. Metallic and dielectric screening, for instance, are fundamentally different physical responses, yet state-of-the-art MLIPs typically incorporate only one of them: charge-equilibration schemes assume a global metallic response and cannot distinguish it from dielectric screening when the latter occurs. Choosing a suitable potential for a specific electrochemical application is correspondingly difficult.

The tests require only single-point evaluations and no production dynamics, and thus form an elementary diagnostic routine alongside traditional energy and force metrics. Because they are qualitative, they define necessary rather than sufficient conditions: passing does not ensure quantitative precision, but failing exposes fundamental limitations of an architecture before it is deployed in production simulations. We expect these benchmarks to help practitioners select MLIPs suited to their research questions, and to give developers physically interpretable targets for the next generation of long-range machine-learned potentials.  

\include{acknowledgements}

\bibliography{bib}
\end{document}


\title{Supplemental Information for 'Electrostatic Phenomenology Benchmarks for Machine‑Learned Potentials in Electrochemistry -- Beyond the Energy‑Force Metric'}

\maketitle

\section{Asymptotic energy for charged slabs in periodic boundary conditions}

Consider the following setup: a dielectric slab of thickness L with dielectric constant $\epsilon_r$
(relative to the (absolute) vacuum permittivity $\epsilon_0$)
with a constant charge density $n$.
Given boundary conditions $V(0)=V(L)=0$ and requiring a  symmetric solution, the potential is
\begin{equation}
V(z) = \frac{z(L-z)n}{2\epsilon_r\epsilon_0}\;.
\end{equation}
The average of this potential across the entire slab is given by
\begin{equation}
\overline V = \frac{1}{L}\int_0^{L}dz ~V(z) = \frac{nL^2}{12}\;.
\end{equation}

Next, consider a system consisting of such a dielectric slab with thickness $s$,
and a vacuum region of thickness $v$. If the total charge on the slab is $Q$ per area $A$, a
compensating homogeneous background for the entire system  is 
\begin{equation}
n_{\rm bg}=-\frac{Q}{A(s+v)}
\end{equation}

The electrostatic energy of the system can be expressed by the electrostatic
energy of the two subsystems considering the homogeneous background density,
or equivalently to the energy of the real charge with the average system potential
aligned to zero. We will take the former route.
For simplicity, let's assume the charge is distributed equally across the slab.
If it is not, the energy difference between the actual
charge distribution and the homogeneous case will asymptotically ($v\rightarrow \infty$) behave like
\begin{equation}
\Delta E = C + \mathcal O\big((s+v)^{-1}\big)\;,
\end{equation}
with a constant $C$. The latter term arises from the interaction with the homogeneous background.

In the homogeneous case, the effective charge density (incl. the background) in the slab
region becomes
\begin{equation}
n_{1} = \frac{Q}{As} + n_{\rm bg} = \frac{Qv}{As(s+v)} = \frac{Q_1}{As}
\end{equation}
with the total effective charge in this region being
\begin{equation}
Q_1=\frac{v}{s+v}Q \;. 
\end{equation}
The total electrostatic energy in this region can be be written as
\begin{equation}
E_1 = \frac 12 \int_0^s dz~n_1 V_1(z) = \frac 12 Q_1 \overline V_1 = \frac{Q^2 s v^2}{24\epsilon_0\epsilon_r A (s+v)^2}
\;.
\end{equation}
Similarly, the energy in the vacuum region (with $n_2=n_{\rm bg}$, $Q_2=-Q_1$, $\epsilon_r = 1$, and $L=v$) becomes
\begin{equation}
E_2 = \frac 12 Q_2 \overline V_2 = \frac{Q^2 v^3}{24 \epsilon_0(s+v)^2}
\;.
\end{equation}

Summing the two contributions together and rearranging terms yields
\begin{eqnarray}
E_1 + E_2 =& \frac{Q^2 v}{24\epsilon_0 A}&\left(\frac{v}{s+v}\right)\left(\frac{s\frac{1}{\epsilon_r}+v}{s+v}\right)
\nonumber\\
=& \frac{Q^2v}{24\epsilon_0 A}&\Big(1 - \frac{s}{s+v}\left[2-\frac{1}{\epsilon_r}\right] 
\nonumber\\&&
+ \frac{s^2}{(s+v)^2}\left[1-\frac{1}{\epsilon_r}\right]\Big)\;.
\end{eqnarray}

Combining this with the asymptotic behaviour of $\Delta E$ due to the actual (non-homogeneous) charge distribution across the dielectric slab yields an asymptotic scaling
\begin{equation}
E_1 + E_2 + \Delta E = C + \frac{Q^2v}{24\epsilon_0 A} + \mathcal O\big((s+v)^{-1}\big)
\end{equation}

\section{Obtaining the dipole moment of a repeated slab system via energies}

The water dipole and Fermi level pinning tests are most easily diagnosed by inspecting at the
dipole moment and charges, respectively. However, these quantities can also be extracted
from energy and force calculations only of auxiliary systems. Of course, this works
only if the auxiliary systems are not suffering from electrostatic artifacts in their
own right. For completeness, we describe the required setups in the following.
Yet, we strongly recommend to use this approach only as a last resort if charges and
potentials are inaccessible.

To measure the dipole moment of a slab system perpendicular to the surface $\mu_z$, we
recall that a dipole induces a potential drop
\begin{equation}
\Delta V = -\frac{\mu_z}{A \epsilon_0}
\;.
\end{equation}

If periodic boundary conditions are applied, and there is no dipole correction, this implicitly
induces a compensating field
\begin{equation}
\mathcal E = \frac{\Delta V}{c} = \frac{\mu}{\Omega}
\end{equation}
relative to the open boundary case, where $c$ is the total length of the periodic cell along
the out-of-plane $z$-direction, and $\Omega$ is the cell volume.

One obvious probe for the field in the vacuum region is therefore to place
a molecule with a known dipole moment in this region, and flip its orientation as
described in the main text. If the magnitude of the probe dipole is too large compared to the
slab dipole of interest, one can exploit the surface area. By replicating
the slab system in the $xy$-direction (explicit supercell), but not the probe dipole,
the surface dipole per area does not change, while the probe-dipole density can be reduced.

Unfortunately, this test requires very accurate energies. An alternative setup is to
use a probe charge of known, constant magnitude $Q_{\rm probe)}$ (see main text how to obtain this value
from a scaling test), and test the potential drop directly. For this, one creates an
artificial symmetric setup with two identical slabs mirrored in the $z$ direction
and sufficient vaccuum to separate them on both sides.
This duplicated symmetric system has -- by construction -- no net dipole, but the two
vacuum regions lie at different potentials. One can now place the probe ion in
either vacuum region, and compute the potential difference from the energy difference via
\begin{equation}
    \Delta V = \frac{\Delta E}{Q_{\rm probe}}
    \;.
\end{equation}
Again, the probe charge density can be reduced by replicating the slab in the $xy$ direction.

The slab dipole test can also be used to test for Fermi level pinning effect.
Instead of focussing on the charge, we recall that Fermi level pinning implies
that the voltage drop between the surface and the adsorbate is insensitive
to (small) displacements. Thus, the test consists in probing for the voltage drop
as described above when the position of the electrically active adsorbate is varied,
while the force on the adsorbate at height $d$ above the surface scales like $1/d$.
Note that in this case, it is crucial that the probe charge density is negligible
compared to the charging effect under investigation, which can be achieved by
replication in the in-plane direction as described above.

\section{Calculations for the image charge effect}

We constructed a Pt (111) 4$\times$4 surface slab with a K atom placed at different distances $z$ from the surface. The first solvation shell of K is included with six water molecules in octahedral coordination. Calculations were done in the range of $z$ = 10 - 15 $\mathrm{\AA}$, with a total cell length in the $z$ direction of 45 $\mathrm{\AA}$. 

Static DFT calculations are performed using the Vienna ab-initio simulation package (VASP) \cite{kresse1993ab,kresse1994ab,kresse1994norm,kresse1999from} with the PBE functional and an electronic convergence criterion of 10$^{-6}$ eV. We use a plane wave cutoff of 600 eV and a \mbox{$1\times 1 \times 1$} $\mathbf k$-point mesh. We perform the calculations with a total net charge of +1, in hope of obtaining a K$^+$ atom and a neutral Pt slab. However, as shown in the main text, the +1 charge does not fully localize on the K atom. The net charge of K as a function of $z$ is calculated by integrating the obtained charge density surrounding the K atom.

Alternatively, we replaced K with a probe point charge. The probe charge is inserted using the VASP-Python interface (details of the method in \cite{doi:10.1021/acs.jctc.5c02150}). In this case, we got rid of the partial charge localization issue and retrieve a linear $E_{\mathrm{tot}}$-$1/z$ dependence.  


\section{Water dipole}
Molecular dynamics (MD) calculations were performed on the Au (111)/water interface system with a total of four Au layers (64 atoms) and 60 water molecules. A layer of Ne counter electrode was added to confine the water slab. The Langevin thermostat was used to perform NVT calculations at $T$ = 400 K, with 20 ps of pre-equilibration and a 20 ps production run. The time step was set to 0.5 fs. Dipole correction was applied in the $z$ direction. We have used the PBE functional with a 500 eV energy cut-off and a \mbox{$1 \times 1 \times 1$} $\mathbf k$-point mesh.  

\section{LO-TO phonons}
We performed static DFT-LDA calculations of
MgO cells elongated in the $[100]$ direction,
namely a $20\times 1\times 1$ of a
suitable 4-atom orthorhombic unit cell. The final cell dimensions were $80.3\,\textrm{\AA} \times 4.15\,\textrm{\AA}/\sqrt 2\times 4.15\,\textrm{\AA}/\sqrt 2$. We used a PAW plane-wave formalism as implemented in the SPHInX code \cite{boeck2011object} with a cutoff energy of 450\,eV and a \mbox{$1\times 4\times 4$} $\mathbf k$-point folding.
The electronic energy 
convergence for direct energy minimization algorithm was set to $2.7\cdot10^{-7}$\,eV ($10^{-8}$ Hartree).
For the TO phonon, a sine-wave modulation of positions along $[010]$ with an amplitude of 0.05\,\AA{} was applied for each atom. For the LO phonon (modulation of $x$-coordinate along $[100]$), the amplitude was reduced to 0.03\,\AA{} to avoid dielectric breakthrough.

\section{Fermi level pinning}
We performed static DFT calculations of a hydrogen atom placed at different distances $z$ away from a gold (111) surface. The PBE functional and a \mbox{$1 \times 1 \times 1$} $\mathbf k$-point mesh was used. Energy cut-off was set to 500 eV, with an electronic convergence criterion of 10$^{-5}$ eV. Dipole correction was applied in the $z$ direction.

To calculate the effective charge of the hydrogen atom $q^{\rm eff}$, we applied a constant electric field between the gold surface and the Ne counter electrode using the computational counter electrode method \cite{Surendralal2018PRL, doi:10.1021/acs.jctc.5c02150}. $q^{\rm eff}$ is calculated as
\begin{equation}
    q^{\rm eff} = (F_z^{\rm field} - F_z^{\rm 0})/\mathcal E_{\rm ext},
\end{equation}
where $F_z^{\rm field}$ and $F_z^{\rm 0}$ are the forces on the H atom in the $z$ direction, and $\mathcal E_{\rm ext}$ is the applied field strength. 
\bibliography{bib}

%% file: acknowledgements.tex
\section{Acknowledgments}
Part of this work was performed while the B.S., R.V., S.K.E., C.S.C., R.H., K.R., M.T., C.F., J.N. were visiting the Institute for Pure and Applied Mathematics (IPAM), which is supported by the National Science Foundation (Grant Nos. DMS-1925919 and DMS-2422832).
This research was sponsored by the U.S. Army Research Office (ARO) and the U.S. Air Force Office of Scientific Research (AFOSR) under Grant No.~FA9550-23-1-0505, ``Unraveling the Role of Cation Solvation in Aqueous Zn-Ion Batteries: A Combined Theoretical and Experimental Approach.'' The views and conclusions contained in this document are those of the authors and should not be interpreted as representing the official policies, either expressed or implied, of the ARO, the AFOSR, or the U.S. Government. The U.S. Government is authorized to reproduce and distribute reprints for Government purposes notwithstanding any copyright notation herein.
S.R. additionally 
acknowledges financial support from the NRF, funded by the Ministry  of Science and ICT (grant no. RS-2025-02317654).
M.T., C.F., and J.N. acknowledge funding by the Deutsche Forschungsgemeinschaft (DFG, German Research Foundation) through SFB1625, Project No. 506711657, as well as through SFB1394, Project No. 409476157, and for support under Germany’s Excellence Strategy — EXC 2033-390677874-RESOLV.